\documentclass[prl,twocolumn,showpacs,epsfig,longeq]{revtex4}
\usepackage{dcolumn}
\usepackage{amssymb}
\usepackage{bm}
\usepackage[pdftex]{graphicx}
\usepackage{psfrag}

\def\3dots{\:\raisebox{-0.5ex}{$\stackrel{\textstyle.}{:}$}\:}
\def\beq{\begin{equation}}
\def\eeq{\end{equation}}
\def\bea{\begin{eqnarray}}
\def\eea{\end{eqnarray}}

\begin{document}

\title{Nonequilibrium Fluctuation Relation for Sheared Micellar Gel in a Jammed State}

%\author{Sayantan Majumdar$^{1}$ and {A.K. Sood}\thanks{To whom correspondence should      
%be addressed}}                                                                   
%\email{asood@physics.iisc.ernet.in} 
%\affiliation{Department of Physics, Indian Institute of Science,
%Bangalore 560012, INDIA}
\author{Sayantan Majumdar} 
\author{A.K. Sood}\thanks{corresponding author}                                                                      
\email{asood@physics.iisc.ernet.in}                                             
\affiliation{Department of Physics, Indian Institute of Science,
Bangalore 560012, INDIA}             

\date{\today}
\draft
\begin{abstract}
We show that the shear rate at a fixed shear stress in a micellar gel in a jammed state exhibits large fluctuations, showing positive and negative values, with the mean shear rate being positive. The resulting probability distribution functions (PDF's) of the global power flux to the system vary from Gaussian to non-Gaussian, depending on the driving stress and in all cases show similar symmetry properties as predicted by Gallavotti-Cohen steady state fluctuation relation.
The fluctuation relation allows us to determine an effective temperature related to the structural constraints of the jammed state. We have measured the stress dependence of the effective temperature. Further, experiments reveal that the effective temperature and the standard deviation of the shear rate fluctuations increase with the decrease of the system size.
 
\end{abstract}

\maketitle
%\begin{multicols}{2}
%\narrowtext
For a system driven arbitrarily far from equilibrium by a large perturbation, the traditional linear response theory and the fluctuation dissipation theorem do not apply. However, remarkably strong Fluctuation Theorems (FT) have been obtained for a variety of driven systems arbitrarily far from equilibrium. Motivated by the molecular dynamics simulation results on sheared hard disks in two dimensions \cite{Evanshardsphere}, Evans and Searles derived a FT for systems going from an equilibrium to a non-equilibrium steady state \cite{Evans} and Gallavotti and Cohen derived FT for non-equilibrium stationary state systems \cite {Gallavotti}. The Gallavotti- Cohen Steady State Fluctuation Relation (SSFR) based on chaotic hypothesis says \cite{Gallavotti},
\begin{equation}
%\begin{centre}
%\pi=1   
Lt_{\tau\rightarrow\infty}P(+s_{\tau})/P(-s_{\tau})= e^{\Sigma\,{\tau}s_{\tau}},
\end{equation}
with $\Sigma$ = 1.
Here, $s_{\tau} = \frac{1}{\tau}\displaystyle\int^{t+\tau}_{t}s(t^{'})\,dt^{'}$ and $s(t)$ is the rate of  entropy production in the non-equilibrium steady state. $P(+s_{\tau})$ is the probability of observing a fluctuation of magnitude $s_{\tau}$ over a phase space trajectory of duration $\tau$ which is larger than any microscopic time scale of the system. Naturally, $P(-s_{\tau})$ gives the probability of transient violation of second law of thermodynamics for the time $\tau$, as the entropy decreases over this time. The physical implication of Eq(1) is that, if the value of $s_{\tau}$ and $\tau$ is large, as in case of macroscopic systems and time scales, $P(+s_{\tau})>>P(-s_{\tau})$, i.e. the probability of observing entropy increasing fluctuations are overwhelmingly large compared to those in which entropy decreases. Thus, in classical thermodynamics we never see the decrease in entropy in any physical process. Extension of steady state fluctuation theorem for finite times is discussed in \cite{vanZon}. 

The experiments on Fluctuation Relation (FR) reported so far can be broadly divided into two classes. Experiments on systems with small number of degrees of freedom include dragging of a Brownian particle in an optical trap \cite{{Carberry},{Wang}}, electrical circuits \cite{Garnier}, RNA stretching \cite{{Collin},{Liphardt}} where RNA free energy between folded and unfolded states were estimated using Crook's Relation and Jarzynski Equality and stochastic harmonic oscillators \cite{{Cilijstat},{Ciliprl}}.   
The second class of experiments which is of relevance here, includes macroscopic systems with large number of degrees of freedom  
such as, Rayleigh-Benard convection \cite{{Cilliberto},{Shang}}, pressure fluctuations on a surface kept in turbulent flows \cite{Cilliphysica}, vertically shaken granular beads \cite{Menon}, Lagrangian turbulence on a free surface \cite{Bandi_turbulence} and liquid crystal electro-convection \cite{electroconvection}. To our knowledge no experimental evidence exists for the FR in large volume sheared fluids. In this Letter we address, for the first time, instance of the FR in case of a macroscopic sized sheared micellar gel in a jammed state. In our context $s(t)=\frac{P(t)}{k_{B}.T_{eff}}$, where P(t) is the instantaneous power flux into the system and $T_{eff}$ is the effective temperature of the system. We show that the nature of PDF's of global power flux for the same system can be Gaussian or non-Gaussian, depending on the applied stress \cite{Gallavotti1}. Further, the PDF's  show similar symmetry properties as predicted in Eq.(1), even in non -trivial cases where PDF's show large deviation from the Gaussian nature. An important point is that in our sheared system, the dominant noise is not thermal but rather athermal, i.e it arises in the very process of driving, as in \cite{Menon}. Our results also provide a method to estimate the effective temperature $T_{eff}$ of the driven jammed state, like in macroscopic granular system \cite{Menon}. 
It should be emphasized that $T_{eff}$ is an effective structural temperature estimated from the non-equilibrium fluctuations resulting from the applied stress and has nothing to do with the actual temperature of the system. We will show that  
$T_{eff}$ increases with the increasing driving stress as well as decreasing system size. This result may have important significance in statistical formulation of driven complex systems.    

Our experiments are carried out on surfactant Cetyltrimethylammonium Tosylate(CTAT)-water system in the concentration range 35 - 41 wt\%  \cite{Soltero} where a hexagonal phase of surfactant cylindrical micelles is formed.
The samples are prepared by dissolving known amount of CTAT (Sigma Aldrich) in double distilled water and are kept for equilibration for a week at $60^{0}$C. All the experiments are carried out at $30^{0}$C using MCR 300 stress controlled rheometer (Anton Paar, Germany) which has a minimum angular resolution of 0.01 $\mu$rad. A humidity chamber is used to minimize the evaporation of water from the sample during the rheology experiments. We have used cone and plate (CP) and parallel plate (PP) geometries with the rheometer. For both of them the bottom plate is fixed and top plate (cone with angle $1.99^{0}$ in case of CP) rotates to apply shear on the sample kept between the plates. For CP the sample volume is $1.3\times10^{-7} m^3$ and for PP the sample volume was varied between $0.98\times10^{-7} m^3$ and $0.49\times10^{-8} m^3$  as the gap is changed from 200 $\mu$m to 10 $\mu$m, respectively. The results presented in Fig.1 and Fig.2 are done in CP and the system size dependence (Fig.3) is done in PP.   
The sample is subjected to a constant shear stress ($\sigma$) and the shear rate ($\stackrel{.}{\gamma}$) is measured as a function of time.
The time resolution is 40 ms between two consecutive data points in the shear rate measurements. 
The shear rate shows interesting time dependence, which is known as 'aging' and 'shear rejuvenation' \cite{Bonn}. Here, below a critical shear stress, the shear rate decreases with time (i.e. viscosity increases with time, known as 'aging') as shown in Fig.1a, whereas at higher stresses, the viscosity decreases with time (known as 'rejuvenation' which is not shown in Fig.1). To characterize the jammed state we carried out a stress sweep on the sample (Fig.1c) with a waiting time of 20s for each data point, after the steady jammed state is fully reached (Fig.1b). We see that below a stress of $\sim$5 Pa the shear rate is very small but starts to increase beyond 5 Pa. Thus, below 5 Pa stress the sample is in a jammed state which acts like a soft solid under small perturbations. The solid line in Fig.1c marks the approximate boundary separating the jammed state and rejuvenated state. We also carried out a frequency sweep measurement on the jammed steady state as shown in Fig.1d by applying a sinusoidal stress of amplitude 1 Pa.
Over the entire frequency range, elastic modulus G' remains larger than viscous modulus G" signifying the solid-like behaviour of the jammed state. The crossover frequency is $<$0.008 rad/s, implying that the relaxation time is $>$120s.
The main focus of this paper is in the region of jamming for CTAT 39 wt\%. For an applied stress of 2 Pa, 
the sample goes into a stress-induced jamming phase after initial aging for $\sim$ 60s, (Fig.1a).
The nature of the shear rate fluctuations when the jammed state is fully reached is shown in Fig.1b, which shows almost equal number of positive and negative values. In the steady state the average shear rate is $<\stackrel{.}{\gamma}> = 3.66\times10^{-5} s^{-1}$. The range of fluctuations are much higher than the instrument resolution \cite{note}.
The global power flux at time t from the rheometer drive to the system, $P(t) = \sigma\stackrel{.}{\gamma}(t)V_s$, where $V_s$ is the volume of the sample.
We define a normalized variable, $W(t) = s(t)/<s(t)> = \stackrel{.}{\gamma}/<\stackrel{.}{\gamma}>$ and $W_{\tau} = \frac{1}{\tau}\displaystyle\int^{t+\tau}_{t}W(t^{'})\,dt^{'}$. Obviously, $W_{\tau} = s_{\tau}/<s(t)>$ where $<s(t)> = \sigma<\stackrel{.}{\gamma}>V_s/k_{B}T_{eff}$ and $<\stackrel{.}{\gamma}>$ is the time averaged value of shear rate in the steady state over entire run time. Taking $\Sigma$ = 1 as in \cite{Menon},
Eq(1) can be simply written in terms of normalized variable, in the large $\tau$ limit as,
\begin{equation}
R = \frac{1}{\tau}.ln[\frac{P(+s_{\tau})}{P(-s_{\tau})}] = \frac{1}{\tau}.ln[\frac{P(+W_{\tau})}{P(-W_{\tau})}]= s_{\tau}  =W_{\tau}.<s(t)>
\end{equation}
 
Using time-series analysis, we see that the data (Fig.1b) do not correspond to low-dimensional chaos \cite{rajesh}. From these fluctuations in the shear rate, we construct $W_{\tau}$ time series. At this point we want to make a brief comment on the averaging procedure. In constructing the $W_{\tau}$ time series, we divided the $s(t)$ series into different bins of length $\tau$. To improve statistical accuracy we have also taken overlapping bins. To ensure independent sampling, the centre of each bin is shifted from the previous one by a time difference (0.12s) larger compared to the correlation time which is $\leq$0.04s. The PDF is the same as obtained by non-overlapping bins, but with higher statistical accuracy. The PDF's for $W_{\tau}$ are strongly non-Gaussian for integrated power flux for all $\tau$'s. We have shown the PDF's for $\tau$ = 0.4s (Fig.1e) and $\tau$ = 0.56s (Fig.1f) and the solid lines correspond to Gaussian fits to the data. In both the cases, PDF's show strong deviations from Gaussian nature for $\left|W_{\tau}\right| >$ 5, but the quantity ln$[P(+W_{\tau})/P(-W_{\tau})]$ goes linearly with $W_{\tau}$, up to $W_{\tau}\sim$ 9 as shown in Fig.1g. The result is non-trivial and correspond to almost one order of magnitude variations in PDF's. Fig.1h shows a plot of 
R vs $W_{\tau}$ where all the curves scale into a single master curve; a straight line passing through the origin, thus agreeing with  Eq.(2). The corresponding slope gives $<s(t)>$ = 0.8$\pm$ 0.006 $s^{-1}$ which corresponds to an effective temperature $T_{eff} = (8.8 \pm 0.07)\times10^{11}$K. 

\begin{figure}[htbp]
\includegraphics[width=0.45\textwidth]{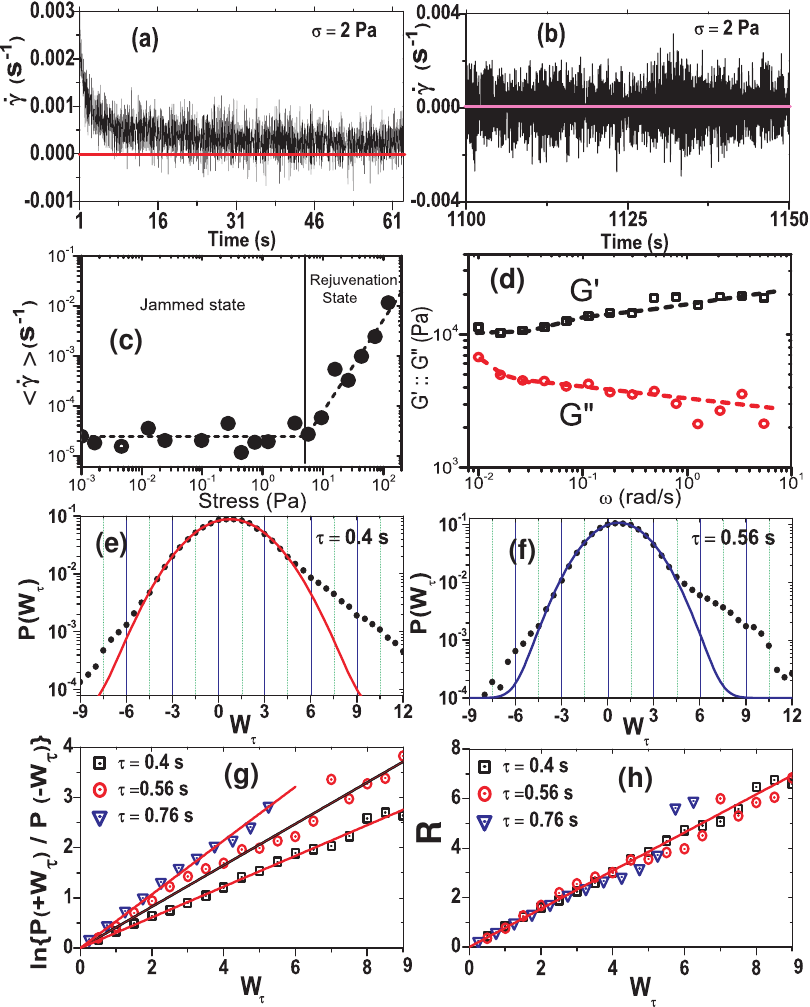}
\caption{(a) The initial aging part for CTAT 39wt\% for the applied stress $\sigma$ = 2 Pa. (b) Typical shear-rate fluctuations for CTAT 39wt\% for $\sigma$ = 2 Pa. (c) Average shear rate ($<\stackrel{.}{\gamma}>$) vs stress ($\sigma$) and (d) Storage (G') and loss (G") modulus vs angular frequency for CTAT 39 wt\% sample (stress amplitude = 1 Pa ). The dashed lines are guide to the eye.
Probability distribution functions of $W_{\tau}$ for (e) $\tau$ = 0.4s and (f) $\tau$ = 0.56s, which deviate strongly from Gaussian nature as seen from the Gaussian fits (solid lines). 
(g) Plot of ln$[P(+W_{\tau})/P(-W_{\tau})]$ vs $W_{\tau}$ for different $\tau$'s: 0.4s, 0.56s, 0.76s against $W_{\tau}$, solid lines are the straight line fits to the data. (h) Plot of 
R = $\frac{1}{\tau}.ln[P(+W_{\tau})/P(-W_{\tau})]$ vs $W_{\tau}$ for different $\tau$'s: 0.4s, 0.56s, 0.76s. Here, all the curves scale into a straight line passing through the origin, as shown by the fitted solid line.} 
\label{Figure2}
\end{figure}

We will now present the result for CTAT 39 wt\%, when the applied stress is reduced to 0.5 Pa as shown in Fig.2.
The experiments are done on fresh samples from the same batch of CTAT (39wt\%). In this case the sample goes to a stress-induced jamming state as soon as the experiment is started. The average shear rate for the steady jammed state is $<\stackrel{.}{\gamma}> = 3.77\times10^{-5} s^{-1}$ which within experimental errors is same as for 2 Pa stress. The typical nature of shear rate fluctuations in the jammed state are shown in Fig.2a. In this case, the probability distribution functions (PDF's), $P(W_{\tau})$ of $W_{\tau}$ (Fig.2b) are perfectly Gaussian for all $\tau$'s (only three are shown in the figure for clarity), as depicted by the Gaussian fits in the figure almost over three orders of magnitude. From these PDF's we again estimate the quantity
R for different $\tau$'s and plot them against $W_{\tau}$, as shown in Fig.2c. All the curves again scale into a straight line passing through origin. The magnitude of the slope of the straight line gives $<s(t)>$ = 1.2 $\pm$ 0.0032 $s^{-1}$ which gives $T_{eff} = (1.5 \pm 0.004)\times10^{11}$K. We have determined $T_{eff}$ for eight different values of the applied shear stress ($\sigma$), shown in log-linear plot, Fig.2d. It can be seen that $T_{eff}$ increases with $\sigma$, similar to the variation of effective temperature with respect to shear rate in \cite{Ono}. We have done experiments for a few more concentrations of CTAT. For samples with a concentrations of 35 wt\%, the nature of the PDF's for $W_{\tau}$ remains non-Gaussian for a stress level of more than 0.5 Pa. The non-Gaussian fluctuations of $W_{\tau}$ are observed at higher stress values as the surfactant concentration is increased.

\begin{figure}[htbp]
\includegraphics[width=0.4\textwidth]{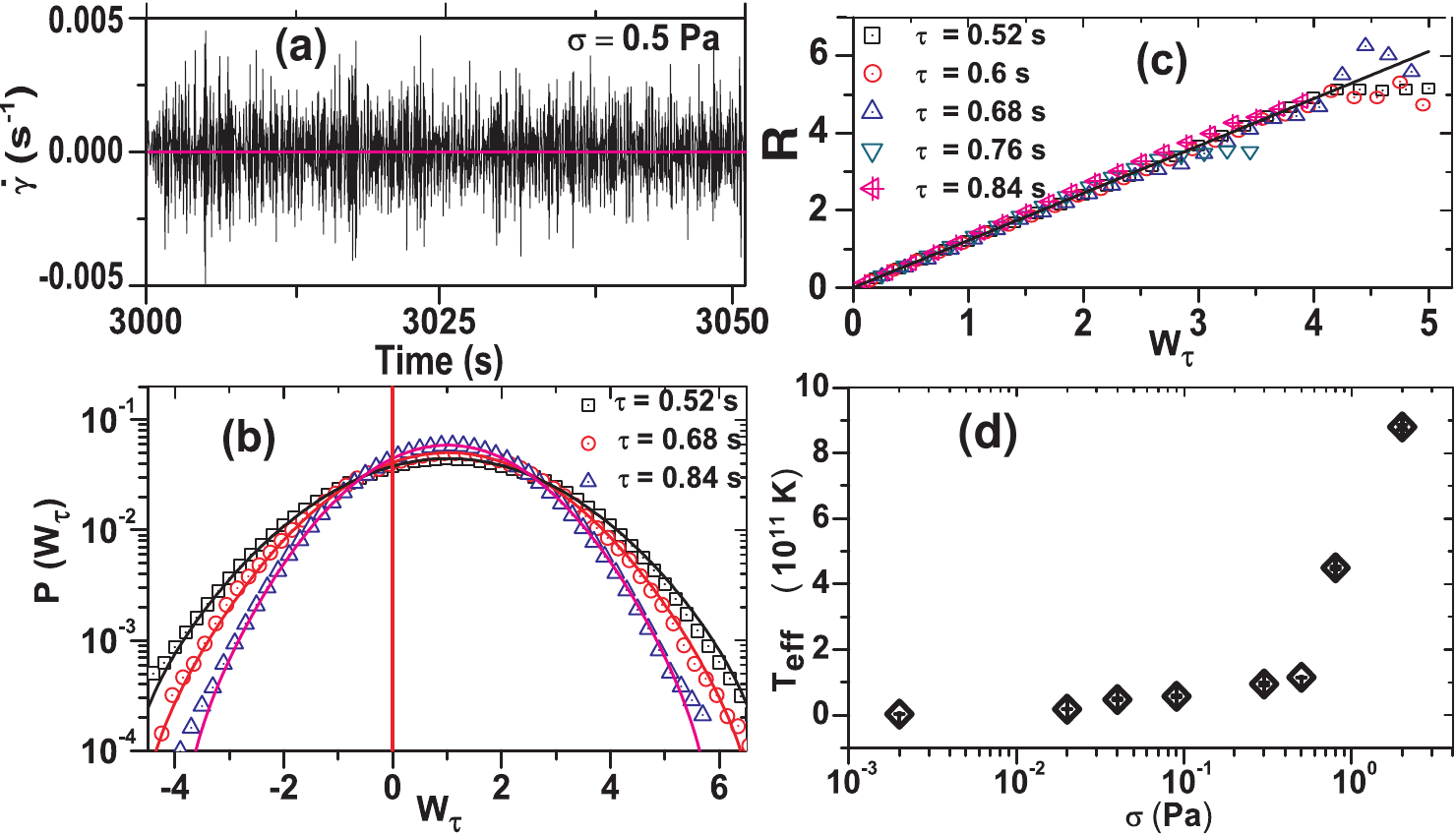}
\caption{(a) Typical shear-rate fluctuations for CTAT 39wt\% for $\sigma$ = 0.5 Pa. (b) Probability distribution functions $P(W_{\tau})$ vs $W_{\tau}$. The solid lines are the fits to the Gaussian function. (c) 
R = $\frac{1}{\tau}.ln[P(+W_{\tau})/P(-W_{\tau})]$ vs $W_{\tau}$ for different $\tau$'s: 0.52s, 0.6s, 0.68s, 0.76s, 0.84s. All the curves scale into a straight line passing through the origin, shown by the fitted solid line. (d) $T_{eff}$ vs applied stress $\sigma$.} 
\label{Figure1}
\end{figure}

We will now turn our attention to the system size dependence of the nature of the observed fluctuations. For this purpose, we have used PP as mentioned earlier, because in CP there is a limit to study system size dependence (when the tip of the cone touches the plate, but still appreciable amount of sample remains in between). This problem can be overcome by using PP where the gap between the plates can be reduced to a arbitrarily small value. We are aware that PP is not ideal for rheological measurements since in this geometry, the shear rate increases linearly with the radial distance from the centre of the plate (which is compensated in CP by adjusting the angle of the cone). We will, therefore, get an effective value of the shear rate. Since we are interested at this point, not in the absolute values of the shear rate but rather in the statistical properties of fluctuations in $\stackrel{.}{\gamma}$, we'll sidestep this issue. 
The experiments were done for 39wt\% sample for the applied stress of 0.5 Pa, with seven gap thicknesses varying between 10 $\mu$m and 200 $\mu$m. The nature of fluctuations in $\stackrel{.}{\gamma}$ are  shown in Fig.3a for three values of gap thickness. It can be seen that the amplitude of fluctuations increases as the gap is reduced. Consequently, the width of distributions of $W_{\tau}$ also increases. For each gap, the probability distribution function of $W_{\tau}$ (only for 0.72s) is shown in Fig.3b, for clarity.
The PDF's are Gaussian for all $\tau$'s in all the three cases. This is shown by the fit of the data to the Gaussian function $P(W_{\tau}) = 1/\sqrt{2\pi\Gamma^{2}}\, e^{-\,(W_{\tau}-<W_{\tau}>)^{2}/2\Gamma^{2}}$ where $<W_{\tau}>$ and $\Gamma^{2}$ are, respectively, the mean and variance of the PDF's of $W_{\tau}$. Here, the Gaussian nature of fluctuations enables us to estimate the ratio of probabilities directly, namely, $ln [P(+W_{\tau})/P(-W_{\tau})] = {(2<W_{\tau}>/\Gamma^{2})W_{\tau}}$. Thus, symmetry like SSFR requires the quantity
$1/W_{\tau}[ln [P(+W_{\tau})/P(-W_{\tau})]] = {(2<W_{\tau}>/\Gamma^{2})}$ to be a linear function of $\tau$. We have estimated the quantity ${(2<W_{\tau}>/\tau\Gamma^{2})}$ for different $\tau$'s from Gaussian fits and plotted them against $\tau$, for three different gap thicknesses, as shown in Fig.3c. In all cases we get straight lines parallel to $\tau$ axis (Fig.3c) in the large $\tau$ limit, as predicted by SSFR. The effective temperature $T_{eff}$, estimated from Eq.(4) for different gap values (L), is shown in Fig.3e. In calculating $T_{eff}$ we need a value of the sample volume $V_{s}$ which is estimated from the geometry of the shear cell. In practice, a small volume of the sample can protrude out of the shear cell which will add error bars in $T_{eff}$, especially for small L. This error is not possible to estimate accurately.  
It can be seen that $T_{eff}$ increases as L decreases. The dotted line a guide to the eye.   
Further, the standard deviation ($\Gamma$) of the shear rate fluctuations shown in Fig.3d, increases as  L decreases. The solid line corresponds to $\Gamma \propto$ 1/L.    

\begin{figure}[tbp]
\includegraphics[width=0.45\textwidth]{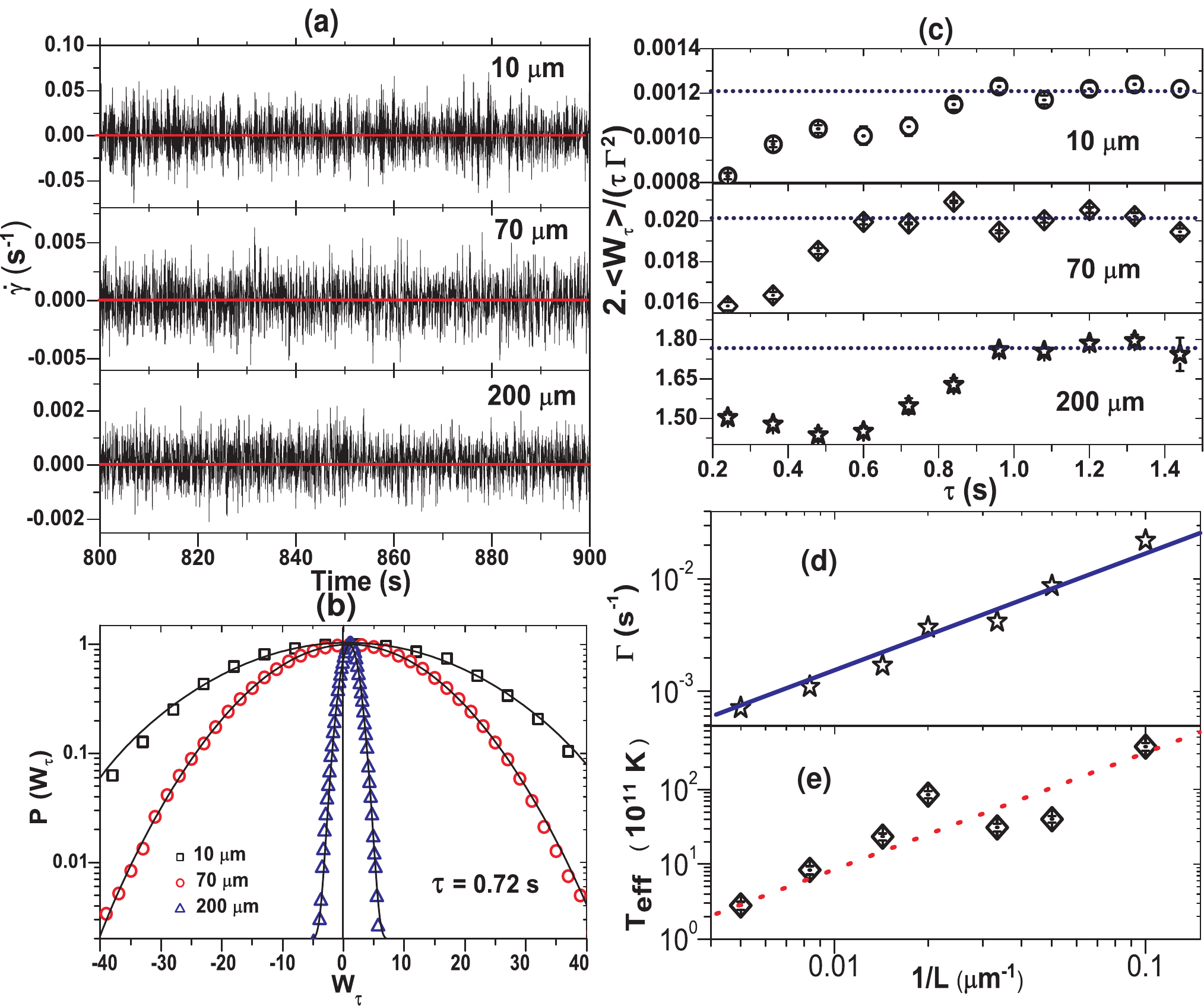}
\caption{(a) The time dependence of $\stackrel{.}{\gamma}$ for different gaps L between the parallel plates (L = 10 $\mu$m, 70 $\mu$m and 200 $\mu$m) for CTAT 39wt\%. (b) Probability distribution functions of $W_{\tau}$ for $\tau$ = 0.72s for L = 10 $\mu$m, 70 $\mu$m and L = 200 $\mu$m. The solid lines are fits to the Gaussian function. (c) $2<W_{\tau}>/\tau\Gamma^{2}$ vs $\tau$. The dotted lines indicate the asymptotic convergence of the function for large $\tau$. (d) Standard deviation $\Gamma$ of shear rate fluctuations as a function of 1/L. The solid line corresponds to $\Gamma \propto$ 1/L. (e) Effective temperature ($T_{eff}$) as a function of 1/L. The error bars are less than the size of the symbol. The dotted line is a guide to eye.}
\label{Figure3}
\end{figure}

To conclude, we have shown the evidence of Fluctuation Relation in a sheared micellar gel in the jammed state. 
The system under study is a macroscopically large system which is expected to display "entropy consuming" fluctuations if the dynamics is determined by long lived temporal and spatial correlations \cite{{Bandi_turbulence},{Gallavotti1}}. These long lived correlations will be present in the jammed state. A physical picture of observing negative shear rate fluctuations is that a rearrangement in the jammed state can result in feeding the elastic energy from the system to the rheometer drive, resulting in negative $\stackrel{.}{\gamma}$ fluctuations.
Remarkably, we could control the nature of PDF (Gaussian or non-Gaussion) for the same system with applied stress as a tunable parameter. 
The symmetry properties of the PDF's like SSFR is verified in the non-trivial case of non-Gaussian fluctuations with good statistical accuracy, due to huge number of large negative fluctuations. We show that an effective temperature of the jammed state can be defined experimentally using FR. It will be interesting to explore the connection between our results and a recent theoretical work on defining the temperature of a static granular assembly \cite{Bulbul}. Our observations opens up a possibility of formulation of statistical mechanics of a driven jammed state in an equivalent stress ensemble. Our preliminary experiments on other systems  such as clay suspensions showing jammed state suggest that the non-equilibrium fluctuations in viscosity, both positive and negative, are generic features of a jammed state. We hope that our experimental observations will stimulate many new experiments and theories on non-equilibrium fluctuations in macroscopic systems. 

AKS acknowledges Council of Scientific and Industrial research (CSIR) of Government of India for support.


\begin{thebibliography}{10}
\providecommand*{\bibinfo}[2]{#2} \providecommand*{\eprint}[1]{#1}
\providecommand*{\url}[1]{#1}

\bibitem{Evanshardsphere}
\bibinfo{author}{D.~J.~Evans}, \bibinfo{author}{E.D.G Cohen}, and \bibinfo{author}{G.~P.~Morris } \bibinfo{journal}{Phys.~Rev.~Lett.} \bibinfo{volume}{\textbf{71}}, \bibinfo{pages}{2401} (\bibinfo{date}{1993}).

\bibitem{Evans}
\bibinfo{author}{D.~J. Evans}, and \bibinfo{author}{D.J Searles}, \bibinfo{journal}{Phys.~Rev.~E.} \bibinfo{volume}{\textbf{50}}, \bibinfo{pages}{1645} (\bibinfo{date}{1994}).

\bibitem{Gallavotti}
\bibinfo{author}{G.~Gallavotti}, and \bibinfo{author}{E.D.G Cohen}, \bibinfo{journal}{Phys.~Rev.~Lett.} \bibinfo{volume}{\textbf{74}}, \bibinfo{pages}{2694} (\bibinfo{date}{1995}); \bibinfo{author}{G.~Gallavotti}, and \bibinfo{author}{E.D.G Cohen}, \bibinfo{journal}{J.~Stat.~Phys.} \bibinfo{volume}{\textbf{80}}, \bibinfo{}{931} (\bibinfo{date}{1995}).

\bibitem{Cilliphysica}
\bibinfo{author}{S.~Ciliberto} {\em et al.}, \bibinfo{journal}{Physica A (Amsterdam).} \bibinfo{volume}{\textbf{340}}, \bibinfo{pages}{240} (\bibinfo{date}{2004}).

\bibitem{Menon}
\bibinfo{author}{K.~Feitosa}, and \bibinfo{author}{N.~Menon}, \bibinfo{journal}{Phys.~Rev.~Lett.} \bibinfo{volume}{\textbf{92}}, \bibinfo{pages}{164301} (\bibinfo{date}{2004}).

\bibitem{vanZon}
\bibinfo{author}{R.~van Zon}, \bibinfo{author}{E.~G.~D.~Cohen}, \bibinfo{journal}{Phys.~Rev.~Lett.} \bibinfo{volume}{\textbf{91}}, \bibinfo{pages}{110601} (\bibinfo{date}{2003}).

\bibitem{Carberry}
\bibinfo{author}{D.~M.~Carberry} {\em et al.}, \bibinfo{journal}{Phys.~Rev.~Lett.} \bibinfo{volume}{\textbf{92}}, \bibinfo{pages}{140601} (\bibinfo{date}{2004}).

\bibitem{Wang}
\bibinfo{author}{G.~M.~Wang} {\em et al.}, \bibinfo{journal}{Phys.~Rev.~Lett.} \bibinfo{volume}{\textbf{89}}, \bibinfo{pages}{050601} (\bibinfo{date}{2002}).

\bibitem{Garnier}
\bibinfo{author}{N.~Garnier}, and \bibinfo{author}{S.~Ciliberto}, \bibinfo{journal}{Phys.~Rev.~E.} \bibinfo{volume}{\textbf{71}}, \bibinfo{pages}{060101} (\bibinfo{date}{2005}).

\bibitem{Collin}
\bibinfo{author}{D.~Collin} {\em et al.}, \bibinfo{journal}{Nature} \bibinfo{volume}{\textbf{437}}, \bibinfo{pages}{231} (\bibinfo{date}{2005}).

\bibitem{Liphardt}
\bibinfo{author}{J.~Liphardt} {\em et al.}, \bibinfo{journal}{Science} \bibinfo{volume}{\textbf{296}}, \bibinfo{pages}{1832} (\bibinfo{date}{2002}).

\bibitem{Cilijstat}
\bibinfo{author}{F.~Douarche}, \bibinfo{author}{S.~Ciliberto}, and \bibinfo{author}{A.~Petrosyan}, \bibinfo{journal}{J.~Stat.~Mech.} (\bibinfo{date}{2005}) P09011.

\bibitem{Ciliprl}
\bibinfo{author}{F.~Douarche} {\em et al.}, \bibinfo{journal}{Phys.~Rev.~Lett.} \bibinfo{volume}{\textbf{97}}, \bibinfo{pages}{140603} (\bibinfo{date}{2006}).

\bibitem{Shang}
\bibinfo{author}{X.~D.~Shang}, \bibinfo{author}{P.~Tong}, and \bibinfo{author}{K.~Q.~Xia}, \bibinfo{journal}{Phys.~Rev.~E.} \bibinfo{volume}{\textbf{72}}, \bibinfo{pages}{015301} (\bibinfo{date}{2005}).

\bibitem{Cilliberto}
\bibinfo{author}{S.~Ciliberto}, and \bibinfo{author}{C.~Laroche}, \bibinfo{journal}{J.~Phys. IV(France)} \bibinfo{volume}{\textbf{8}}, \bibinfo{pages}{215} (\bibinfo{date}{1998}).

\bibitem{Bandi_turbulence}
\bibinfo{author}{M.~M.~Bandi} {\em et al.} \textit{arXiv:nlin.CD/0607037v2}, (2007) and references therein.

\bibitem{electroconvection}
\bibinfo{author}{W. I. Goldburg} {\em et al.}, \bibinfo{journal}{Phys.~Rev.~Lett.} \bibinfo{volume}{\textbf{87}}, \bibinfo{pages}{245502} (\bibinfo{date}{2001}).

\bibitem{Gallavotti1}
\bibinfo{author}{G.~ Gallavotti}, \bibinfo{journal}{Eur.~Phys.~J. B} \bibinfo{volume}{\textbf{61}}, (\bibinfo{date}{2008}), \bibinfo{pages}{1} .

\bibitem{Soltero}
\bibinfo{author}{J.~F.~A.~Soltero}, and \bibinfo{author}{J.~E.~Puig}, \bibinfo{journal}{Langmuir} \bibinfo{volume}{\textbf{11}}, \bibinfo{pages}{3337} (\bibinfo{date}{1995}).

\bibitem{Bonn}
\bibinfo{author}{P.~Coussot} {\em et al.}, \bibinfo{journal}{J.~Rheol} \bibinfo{volume}{\textbf{46}}, \bibinfo{pages}{573} (\bibinfo{date}{2002}).

\bibitem{note}
A typical shear rate fluctuation of 0.001 $s^{-1}$ observed in our experiments and a sample time of 40 ms imply a shear deformation ${\gamma} = 4\times10^{-5}$. This corresponds to an angular displacement $\phi = \gamma \times$ cone-angle ($\beta$) = 1.4 $\mu$rad, which is much larger than the angular resolution of the instrument (0.01 $\mu$rad) which can be found in the technical specifications of the MCR300 rheometer at the site http://www.oleinitec.fi/pdf/Physica/mcr$_{-}$serie$_{-}$e.pdf.
                                                                                            
\bibitem{rajesh}
\bibinfo{author}{R.~Ganapathy} {\em et al.}, \bibinfo{journal}{Phys.~Rev.~Lett.} \bibinfo{volume}{\textbf{96}}, \bibinfo{pages}{108301} (\bibinfo{date}{2006}); \bibinfo{author}{R.~Bandyopadhyay} {\em et al.}, \bibinfo{journal}{Phys.~Rev.~Lett.} \bibinfo{volume}{\textbf{84}}, \bibinfo{pages}{2022} (\bibinfo{date}{2000}).

\bibitem{Ono}
\bibinfo{author}{I.~K.~Ono} {\em et al.}, \bibinfo{journal}{Phys.~Rev.~Lett.} \bibinfo{volume}{\textbf{89}}, \bibinfo{pages}{095703} (\bibinfo{date}{2002}).

\bibitem{Bulbul}
\bibinfo{author}{S.~Henkes} {\em et al.} \bibinfo{journal}{Phys.~Rev.~Lett.} \bibinfo{volume}{\textbf{99}}, \bibinfo{pages}{038002} (\bibinfo{date}{2007}). 


\end{thebibliography}
\end{document}